\documentclass[aps,prd,twocolumn,preprintnumbers,superscriptaddress,dblfloatfix,nofootinbib]{revtex4-1}
\usepackage[utf8]{inputenc}
\usepackage{lmodern}
\usepackage{amsmath,amssymb}
\usepackage{graphicx}
\usepackage{url}
\usepackage{color} 
\usepackage{enumitem}
\usepackage{subfigure}
\usepackage[dvipsnames]{xcolor}
\usepackage[colorlinks=true,breaklinks=true]{hyperref}
\hypersetup{allcolors=[rgb]{0.0 0.0 0.6},linkcolor=[rgb]{0.75 0.05 0.05}}
\usepackage{bm}
\usepackage{epsfig}
\usepackage{slashed}
\usepackage{color}
\usepackage{accents}
\usepackage[dvipsnames]{xcolor}
\usepackage[colorlinks=true,breaklinks=true]{hyperref}
\hypersetup{allcolors=[rgb]{0.0 0.0 0.6},linkcolor=[rgb]{0.75 0.05 0.05}}

\renewcommand\[{\left[}

\newcommand{\exclude}[1]{}

\begin{document}
\preprint{IPMU23-0026} 

\title{$G$ Objects and primordial black holes}

\author{Marcos M. Flores}
\affiliation{Department of Physics and Astronomy, University of California, Los Angeles  (UCLA) \\
Los Angeles, CA 90095}

\author{Alexander Kusenko}
\affiliation{Department of Physics and Astronomy, University of California, Los Angeles  (UCLA) \\
Los Angeles, CA 90095}
\affiliation{
Kavli Institute for the Physics and Mathematics of the Universe (WPI)\\
The University of Tokyo Institutes for Advanced Study, The University of Tokyo\\ Kashiwa, Chiba 277-8583, Japan
}
\affiliation{Theoretical Physics Department, CERN, 1211 Geneva 23, Switzerland}

\author{Andrea M. Ghez}
\affiliation{Department of Physics and Astronomy, University of California, Los Angeles  (UCLA) \\
Los Angeles, CA 90095}

\author{Smadar Naoz}
\affiliation{Department of Physics and Astronomy, University of California, Los Angeles  (UCLA) \\
Los Angeles, CA 90095}
\affiliation{Mani L. Bhaumik Institute for Theoretical Physics\\
Department of Physics and Astronomy, UCLA\\
Los Angeles, CA 90095, USA}	

\date{\today}
	
\begin{abstract}
We suggest that ``$G$ objects" recently discovered in the Galactic Center may be clouds of gas bound by the gravitational field of stellar-mass black holes produced in the interactions of sublunar primordial black holes with neutron stars.  If dark matter is composed of primordial black holes with masses $(10^{-16} - 10^{-10}) M_\odot$, these black holes can be captured by neutron stars in the Galactic Center, where the dark matter density is high. After the capture, the neutron star is consumed by the black hole, resulting in a population of  $(1-2) M_\odot $ black holes. These stellar-mass black holes, accompanied by gaseous atmospheres, can account for the observed properties of the $G$ objects, including their resilience to tidal disruption by the supermassive black hole in the Galactic Center while also producing emission consistent with inferred luminosities.
\end{abstract}
\maketitle
	

\section{Introduction}

The existence of black holes formed in the early Universe remains an open question. These primordial black holes (PBHs) are an attractive candidate for dark matter (DM)~\cite{Zeldovich:1967,Hawking:1971ei,Carr:1974nx,Khlopov:1985jw,Dolgov:1992pu,Yokoyama:1995ex,Wright:1995bi,GarciaBellido:1996qt,Kawasaki:1997ju,Green:2004wb,Khlopov:2008qy,Carr:2009jm,Frampton:2010sw,Kawasaki:2016pql,Carr:2016drx,Inomata:2016rbd,Pi:2017gih,Inomata:2017okj,Garcia-Bellido:2017aan,Georg:2017mqk,Inomata:2017vxo,Kocsis:2017yty,Ando:2017veq,Cotner:2016cvr,Cotner:2017tir,Cotner:2018vug,Sasaki:2018dmp,Carr:2018rid,Germani:2018jgr,Banik:2018tyb,Escriva:2019phb,Germani:2019zez,1939PCPS...35..405H,Cotner:2019ykd,Kusenko:2020pcg,deFreitasPacheco:2020wdg,Takhistov:2020vxs,Biagetti:2021eep}, but also could be responsible for various astrophysical phenomena including seeding supermassive black holes (SMBHs)~\cite{Bean:2002kx,Kawasaki:2012kn,Clesse:2015wea} and could play a role in the synthesis of heavy elements~\cite{Fuller:2017uyd,Takhistov:2017nmt,Takhistov:2017bpt}. Unlike PBHs, astrophysical neutron stars (NSs) and black holes are remnants of core-collapse supernovae, which follow the gravitational collapse of a massive star. The masses of black holes formed in a core-collapse supernova are expected to be above $\sim 2.5\ M_\odot$, while NS masses are expected to be $1.5$ - $2.5\ M_\odot$~\cite{Godzieba:2021MaxNS}. Black holes with masses 5 - 10 $M_\odot$ have been observed in low-mass x-ray binaries, while none have been detected in the first ``mass gap" of $\sim 3$ - $5\ M_\odot$~\cite{Ozel:2010BHDist}. While black holes within the mass gap can have stellar origins~\cite{Ozel:2010BHDist}, detection of black holes below $\sim 3\ M_\odot$ would point to an origin other than stellar evolution. 

Black holes below $3\ M_\odot$, which we shall refer to as solar-mass black holes throughout, can have two different primordial origins. First, various extensions of Standard Model physics are able to produce solar-mass PBHs outright~\cite{Green:2020jor,Garcia-Bellido:1996mdl,Cotner:2016cvr,Cotner:2019ykd,Flores:2020drq}. Though these PBHs can only account for $\lesssim 1 \%$ of DM, they may act as progenitors for gravitational wave events~\cite{Abbott:2016blz,Abbott:2016nmj,Abbott:2017vtc,Clesse:2016vqa,Bird:2016dcv,Sasaki:2016jop}. Second, any population of sublunar-mass PBHs can give rise to a population of 1 - 2 $M_\odot $ black holes through collisions and capture by NSs~\cite{Capela:2013CapNS,Kouvaris:2013kra,Fuller:2017uyd,Takhistov:2017bpt, Abramowicz:2017zbp,Takhistov:2020vxs, Abramowicz:2022mwb}. After a period of accretion, the original low-mass PBH converts its host NS into a solar-mass black hole. These conversion events are most likely to occur in regions of high DM density, particularly the Galactic Center (GC). 

The central parsec of the Milky Way offers a unique probe of whether PBHs make up all or a significant fraction of DM, i.e., the PBH-DM hypothesis. Recently, a population of unusual objects have been found closely orbiting Sagittarius A$^*$ (Sgr A$^*$)~\cite{Ciurlo:2020GObjects}. Located in the central 0.04 pc of the GC, these so-called $G$ objects act dynamically as stellar-mass objects, while simultaneously  showing both thermal dust emission and line emission from ionized gas. Additionally, objects $G$1 and $G$2 have shown an unexpected resilience to tidal disruption.  After passing through their periapse both $G$1 and $G$2 remained intact, even though both experienced tidal interactions during their passage~\cite{Witzel:2017Post, Gillessen:2019G2Drag}. This seems to suggest that the $G$ objects contain a stellar-mass core cloaked in an envelope of gas and dust.

Numerous models have been proposed to explain the origin of $G$1 and $G$2. The lack of tidal disruption at periapse has led authors to consider these objects to be an optically thick distribution of gas and dust surrounding a star. This central star may be a young, low-mass star that has retained a protoplanetary disk~\cite{Murray:2012Proto} or that generated a mass-loss envelope~\cite{Scoville:2013Mass-LossEnv}. $G$ objects may also originate as the merger product of a binary system~\cite{Prodan:2015Secular, Stephan:2016BinaryGOs, Stephan:2019fbg}. For completeness, we also acknowledge that stellar core hypothesis is not universally accepted and other models, e.g., ~\cite{Guillochon:2014bca}, offer possible alternative explanations for $G$2's origin.

Here we propose that $G$ objects may be solar-mass black holes enshrouded by a thick atmosphere of gas. PBHs with sublunar masses can be captured by NSs, transforming the host star into a $1$ - $ 2\ M_\odot$  black hole. This process can contribute to the paucity of NS in the GC, known as the ``missing pulsar problem"~\cite{Dexter:2014GCPulsProb}. The resulting population of black holes have deep potential wells that can retain the gas and dust ejected during the conversion process and  therefore appear as the observed $G$ objects.

In Sec.~\ref{sec:FormationG-Objects} we will discuss the formation of solar-mass black holes from sublunar-mass PBHs and NSs, as well as and the ability of these conversion events to appear as the present population of $G$ objects. In Sec.~\ref{sec:NSCalcs}, we estimated the expected number of converted NSs and demonstrate that this is consistent with the observed population of $G$ objects. In Sec.~\ref{sec:AccretionEstimates} we discuss the emission due to accretion in our scenario, and compare these estimates with observation. Finally, in Sec.~\ref{sec:DiscConclusion} we summarize our scenario, and discuss prospects for future detection and observations. 

\section{Formation of $G$ objects from NSs and PBHs}
\label{sec:FormationG-Objects}

To account for all of DM, PBHs must have masses in the $10^{17}$ - $10^{23}\  g$~\cite{Carr:2020gox} range, and some theories, e.g., supersymmetry, naturally favor this range~\cite{Cotner:2016cvr, Cotner:2019ykd, Flores:2021jas}. In DM rich regions, the average NS lifetime is the sum of three components: the black hole capture time $t_{\rm cap}$, the settling time $t_{\rm set}$, and the time required for conversion of the NS into a solar-mass black hole $t_{\rm acc}$. Additionally, we define the consumption time $t_{\rm con}\equiv t_{\rm set} + t_{\rm acc}$, which will be important for our later discussion.

The NS-PBH capture rate\footnote{Ref.~\cite{Capela:2013CapNS} discussed a possible limit on PBH dark matter if globular clusters contained a significant overdensity of dark matter. However, observational data~\cite{Bradford:2011aq,Ibata:2012eq} constrain the dark matter content in such systems to several orders of magnitude below what is needed for an exclusion limit. } is given by~\cite{Capela:2013CapNS}
\begin{equation}
F \equiv
\frac{\Omega_{\rm PBH}}{\Omega_{\rm DM}}F_0^{\rm MW}
\end{equation}
where $\Omega_{\rm PBH}$ and $\Omega_{\rm DM}$ are the present-day normalized energy density fractions $\Omega_{X} = \rho_X/\rho_{\rm crit}$, superscript MW refers to the Milky Way, and
\begin{eqnarray}
F_0^{\rm MW}
=
\sqrt{6\pi}
\frac{\rho_{\rm DM}}{M_{\rm PBH}\bar{v}}
\left(
\frac{2GM_{\rm NS}R_{\rm NS}}{1 - 2GM_{\rm NS}/R_{\rm NS}}
\right)\\[0.25cm]
\times
\left(
1 - e^{-3E_{\rm loss}/(M_{\rm PBH}\bar{v}^2)}
\right)\nonumber
\end{eqnarray}
is the capture rate assuming PBHs comprise all of DM. Here, $M_{\rm NS}$ is the NS mass, $R_{\rm NS}$ is the NS radius, $M_{\rm PBH}$ is the mass of the interacting PBH, $\bar{v}$ is the PBH velocity dispersion, and $\rho_{\rm DM}$ is the PBH contribution to the DM density. Finally~\cite{Capela:2013CapNS},
\begin{equation}
E_{\rm loss}
\approx
58.8\frac{G^2 M_{\rm PBH}^2 M_{\rm NS}}{R_{\rm NS}^2}
\end{equation}
is the average interaction loss energy during a NS-PBH interaction. The capture rate can be enhanced by considering the velocity dispersion of NSs in a DM rich environment, but for the parameter space we consider this effect will be small. For our analysis, we consider NS with masses $M_{\rm NS} = 1.5\ M_\odot$ and radii $R_{\rm NS} = 10$ km. 

The DM profiles are often modeled with two-step power-law functions of the form
\begin{equation}
\rho_{\rm DM}(r)
\simeq
\frac{\rho_\odot}{(r/r_\odot)^\alpha(1 + r/r_\odot)^{\beta - \alpha}},
\end{equation}
where $\rho_\odot$ is the local DM density for $r_\odot = 8.2$ kpc, $\rho_\odot = 0.4$ GeV cm$^{-3}$~\cite{2017MNRAS.465...76M}. The Navarro–Frenk–White profile corresponds to $\alpha = 1$, $\beta = 3$~\cite{Navarro:1995iw}.
Given that we are working well within the GC, $r/r_\odot\ll 1$, allowing us to simply use
\begin{equation}
\rho_{\rm DM}(r)
\simeq
\rho_\odot
\left(
\frac{r}{r_\odot}
\right)^{-\alpha}.
\end{equation}
While corelike profiles for DM are possible, they tend to be favored in self-interacting or warm DM models~\cite{Tulin:2017ara, Viel:2013fqw}. PBHs instead act as cold, collisionless DM, and, based on the $N$-body simulations without baryons, one expects it to have a cuspy profile~\cite{Navarro:1995iw}. Interactions with the baryonic matter could potentially make the density profile more corelike, although there is also a possibility that baryonic contraction may cause a local enhancement of the DM density near the central SMBH.
To accommodate a variety of models, as well as physical effects such as  adiabatic contraction~\cite{Gnedin:2004cx}, we will take $\alpha\sim 1 - 2$. In addition to this specification, the velocity dispersion $\bar{v}$ is taken to be
\begin{equation}
\bar{v}\to\sigma(r)
=
\sqrt{\frac{GM_{\rm SMBH}}{r(1 + \alpha)}},	
\end{equation}
where $M_{\rm SMBH}\sim 4\times 10^6\ M_\odot$. For example, for $M_{\rm PBH} = 10^{19}\ g$ , we find
\begin{equation}
t_{\rm cap}(r\geq 10^{-2}\ {\rm pc})
\equiv
F^{-1}
\lesssim
\left\{
\begin{array}{lr}
10^{12}\ {\rm yrs}, & \text{for } \alpha = 1,\\
10^{6}\ {\rm yrs}, & \text{for } \alpha = 2,
\end{array}
\right.
\end{equation}
where we assumed that $\Omega_{\rm PBH}/\Omega_{\rm DM} = 1$.

For a typical NS, the time required for a gravitationally captured PBH to settle into the NS core is~\cite{Capela:2013CapNS}
\begin{equation}
t_{\rm set}
\simeq 1.3\times 10^{9}\ \text{yr}\ 
\left(
\frac{M_{\rm PBH}}{10^{19}\ {\rm g}}
\right)^{-3/2}
.
\end{equation}

The final contribution to the the conversion timescale is an estimation of the accretion rate. For simplicity, we will assume spherical Bondi accretion,
\begin{equation}
\dot{M}_{\rm PBH} = 4\pi\lambda_s G^2M_{\rm PBH}^2\rho_{\rm NS}/v_s^3,
\end{equation}
where $M_{\rm PBH}$ is the time evolving mass of the growing black hole, $v_s$ is the sound speed, $\rho_{\rm NS}$ is the NS density, and $\lambda_s$ is an order one parameter depending on the NS equation of state. For a NS described by an $n = 3$ polytrope, $v_s = 0.17$, $\rho_{\rm NS} = 10^{15}$ g cm$^{-3}$, and $\lambda_s = 0.707$~\cite{Shapiro:2008CO}, we find
\begin{equation}
t_{\rm acc}
\equiv
\frac{M_{\rm PBH}}{dM_{\rm PBH}/dt}
\simeq
10\ {\rm yr}
\left(
\frac{10^{19}\ {\rm g}}{M_{\rm PBH}}
\right)
.
\end{equation}
Though we are examining incredibly small black holes, the traditional Bondi accretion rate is still reliable for the PBH-DM mass range~\cite{Giffin:2021kgb}. 

From the three above timescales, we see that the settling timescale determines how quickly a captured PBH will consume its host star, i.e., $t_{\rm con}\simeq t_{\rm set}$. Of these contributions, either $t_{\rm cap}$ or $t_{\rm set}$ dominates the average NS lifetime $\langle t_{\rm NS} \rangle$, depending on the values of $\alpha$ and $M_{\rm PBH}$. Given the average NS lifetime, we wish to determine the number of surviving NSs in a DM rich environment such as the GC. To do so, we model the number of remaining NSs by assuming that the population follows a Poisson process with a decay rate given by 1/ $\langle t_{\rm NS} \rangle$. This allows us to 
define the present-day converted fraction of NSs as the ratio
\begin{equation}
\label{eq:ConvertFrac}
\Upsilon
\equiv
\frac{N_{\rm NS\to BH}}{N_{\rm NS, 0}}
=
1 - \exp(-t_{\rm MW}/\langle t_{\rm NS} \rangle)
\end{equation}
where $t_{\rm MW}$ is the lifetime of the Milky Way galaxy $t_{\rm MW}\sim 1.3\times 10^{10}$ yr.

Again, to account for all of DM, PBHs should have masses in the range of $10^{17}\ g \lesssim M_{\rm PBH} \lesssim 10^{23}\ g$. For masses $10^{20}$ - $10^{23}\ g$, a captured PBH will consume its host NS $\lesssim 10^6$ - $10^8$ yr. Once captured, these larger DM PBHs will convert their host  within the average lifetime of a typical pulsar. For PBHs in this heavier mass range, $10^{6} \lesssim \left\langle t_{\rm NS} \right\rangle \lesssim 10^{12}$ yr for $\alpha$ between 1 and 2, implying that $\mathcal{O}(1 - 100)\%$ of NSs in the GC should have been consumed by PBHs in the age of the Galaxy. This is consistent with the observed underabundance of pulsars in the GC~\cite{Dexter:2014GCPulsProb}. For larger $\langle t_{\rm NS}\rangle$, this result is also consistent with the recent observation of a young, $4\times 10^{4}$-yr-old magnetar J1745-2900~\cite{Mori2013:MagGC,Kennea:2013MagGC} since the magnetar's age is significantly shorter than $\left\langle t_{\rm NS} \right\rangle$. Lighter PBHs, with masses between $10^{17}$ and $10^{20}\ g$, will still convert pulsars but will not do so before the emission of radio waves and eventual spin-down of the host NS. While this lower mass range cannot help in addressing the missing pulsar problem, it still can explain the origin of $G$ objects.

The conversion of a NS into a solar-mass black hole is an extremely violent process. Spin-up occurs as the NS undergoes collapse into the central PBH. Differential rotation can occur as the fractional change in radius is greater for accreted matter in the innermost regions than it is further from the center of the NS. Mass is ejected when matter at the equator reaches the escape velocity. The amount of ejected material may be estimated analytically, though a complete analysis must take into account general relativistic effects~\cite{Owen:1998xg, Andersson:2000mf}.  For the pulsars with the shortest period theoretically possible, i.e., $P = 0.7$ ms, the ejected mass is expected to be of order $0.1\ M_\odot$~\cite{Fuller:2017uyd}. Naturally, longer rotation periods would eject even less mass into the surrounding environment.

Regardless of the amount, the material ejected will occupy the environment closely surrounding the converted NS. The deep potential well of the newly formed solar-mass black hole will allow for the accumulation of the ejected material, forming an atmosphere of dust and gas. We interpret this cloud of material around a central solar-mass black hole as a $G$ object.

Two factors are crucial to the viability of our scenario. First, we must ensure that the number of conversion events within the GC is consistent with the observed number of $G$ objects. Second, we must guarantee that emission predicted from our scenario does not over- or underproduce radiation in a way that is inconsistent with the luminosities inferred by the observation of the known population $G$ objects. The following sections will demonstrate the viability of our scenario across a wide region of parameter space.

\begin{figure}[htb]
    \includegraphics[width=0.95\linewidth]{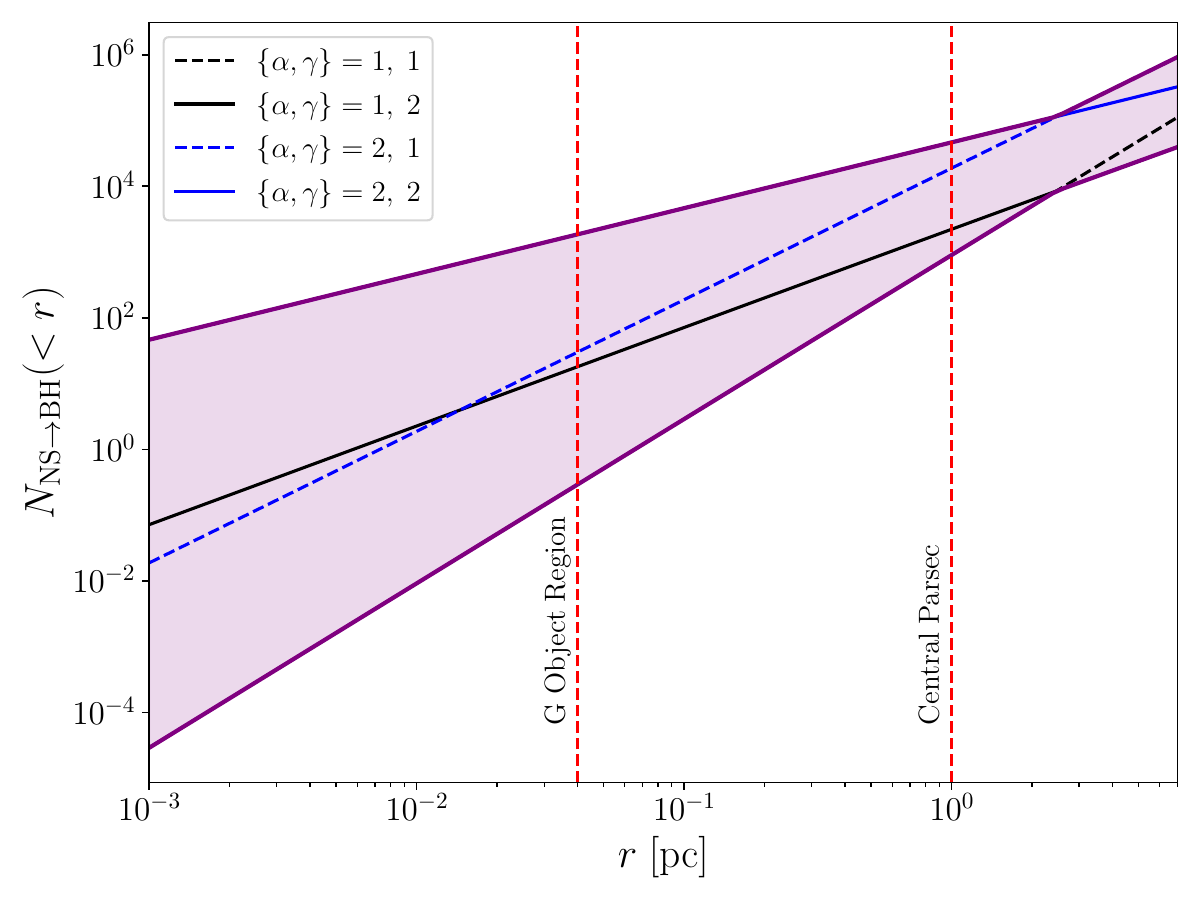}
    \caption{The expected number of NSs converted into black holes  within a volume of radius $r$. The purple shaded region encapsulates varying $\{\alpha,\gamma\}$ between 1 and 2 for each parameter individually. Here we assumed that $M_{\rm PBH} = 10^{21}\ g$.}
	\label{fig:NNStoBH}
\end{figure} 

\section{Numerical Estimates of Converted NSs}

\label{sec:NSCalcs}
The number of converted NSs in the presence of PBH-DM naturally relies on the number of NSs present in the central regions of the Galaxy. To determine the number of NSs within the GC, we utilize results first obtained by Bahcall and Wolf~\cite{Bahcall:1976aa}. In that foundational work, Bahcall and Wolf predicted the distribution of stars around a central SMBH by solving a Fokker-Planck equation for the stellar distribution function. Famously, Ref.~\cite{Bahcall:1976aa} predicted the number density of stars around a central SMBH follows $n(r)\propto r^{-7/4}$, though observations around Sgr A$^*$ suggest a shallower cusp~\cite{Alexander:1999yz, Genzel:2003cn, Do:2009md, Schodel:2007er}. Numerous studies~\cite{Hopman:2006rt, Hopman:2006xn, Keshet:2009vu, Aharon:2016kil,  2022ApJ...940..101L} have also generalized the formalism of Ref.~\cite{Bahcall:1976aa} and depending on their underlying assumptions, obtain different power-law relations for the stellar number density. To account for both theoretical and observational uncertainties, we take the NS number density to be of the form
\begin{equation}
n_{\rm NS}(r) = 
n_{\rm NS,0}
\left(
\frac{r}{r_{\rm NS}}
\right)^{-\gamma},
\quad
\gamma \sim 1 - 2,
\end{equation}
which is consistent with solutions of the Fokker-Planck equation obtained in Refs.~\cite{Bahcall:1976aa, Aharon:2016kil, Hopman:2006rt, Keshet:2009vu, Hopman:2006xn} and observations~\cite{Alexander:1999yz, Genzel:2003cn, Do:2009md, Schodel:2007er}. The normalizations $\{r_{\rm NS}, n_{\rm NS,0}\}$ were determined using the $M-\sigma$ relation~\cite{Tremaine:2002js} such that
\begin{equation}
r_{\rm NS}
=
\frac{G\sqrt{M_{\rm SMBH} M_0}}{\sigma_0^2},
\quad
n_{\rm NS,0}
=
\xi_s r_{\rm NS}^{-3}
\left(
\frac{3 - \gamma}{2\pi}
\right),
\end{equation}
where, in this case, $M_{\rm SMBH} \sim 4\times 10^6\ M_\odot$, $M_0 = 1.3\times 10^8\ M_\odot$, $\sigma_0 = 200$ km sec$^{-1}$, and $\xi_s = 0.014$, determined by the fraction ratios of the populations 
Main Sequence:White Dwarf:NS = 0.72:0.26:0.014
as are typical for continuously star-forming population regions~\cite{2001MNRAS.322..231K}.

The number of NSs within a volume of radius $r$ is given by
\begin{equation}
\label{eq:NumNS}
N_{\rm NS}(r <)
\equiv
4\pi 
\int_0^r
n_{\rm NS}(r')\ r'^2 d r',
\end{equation}
which can be explicitly integrated using the above density leading to
\begin{equation}
\label{eq:IntNumNS}
N_{\rm NS}(r <)
= 
\frac{4\pi r_{\rm NS}^3}{3 - \gamma}\ n_{\rm NS,0}
\left(
\frac{r}{r_{\rm NS}}
\right)^{3-\gamma}	
.
\end{equation}
We find that the number of NSs in the central parsec is $N_{\rm NS}(r < 1\ \text{pc})\simeq \mathcal{O}(10^5)$.
For the region relevant to the observed population of $G$ objects, namely, the inner 0.04 pc,
\begin{equation}
N_{\rm NS}(r < 0.04\ \text{pc}) 
\simeq 
\mathcal{O}(10^1 - 10^3).
\end{equation}
With this, we can finally calculate the expected number of NSs to black hole conversions. This quantity is simply defined as the product of Eqs.~\eqref{eq:ConvertFrac} and~\eqref{eq:IntNumNS}, 
\begin{equation}
N_{\rm NS\to BH}(r < )
\equiv
\Upsilon\cdot N_{\rm NS}(r <)
.
\end{equation} 
The number of converted NSs for various values of $\{\alpha, \gamma\}$ are shown in Fig.~\ref{fig:NNStoBH} with $M_{\rm PBH} = 10^{21} g$ . In this figure, the purple shaded region encloses the possible values of $N_{\rm NS\to BH}(r <)$ for varying DM and NS density profiles. From here we see that, within the region relevant to $G$ objects, various values of $\{\alpha,\gamma\}$ can accommodate the $\mathcal{O}(10)$ objects observed. While enhancements to the DM density within the inner regions of the GC may decrease the average lifetime of a NS within this environment, the limiting factor in determining the number of converted NSs is the NS density within these central regions. That being said, our calculations demonstrate that an enhanced DM density profile \cite{1976ApJ...209..214B, Miralda-Escude:2000kqv} is favorable for generating the $\mathcal{O}(10)$ converted NSs necessary to explain the population of $G$ objects.

\section{Emission of accreting solar-mass black hole}

\label{sec:AccretionEstimates}

It is essential that the emission resulting from accretion of gas and dust onto a central, solar-mass black hole is consistent with observed luminosities of the population of $G$ objects. We will again consider spherical Bondi accretion, which allows us to parametrize the accretion rate in terms of the gas density, gas temperature, and mass of the central black hole~\cite{Shapiro:2008CO},
\begin{equation}
\dot{M}_B
=
2.40\times 10^{10}\ \frac{\rm g}{\rm s}
\left(
\frac{M}{M_\odot}
\right)^2
\left(
\frac{n_\infty}{1\ \text{cm}^{-3}}
\right)
\left(
\frac{T_\infty}{10^4\ \text{K}}
\right)^{-3/2},
\end{equation}
where $M$ is the central black hole mass, $n_\infty$ is the number density of the gas, and $T_\infty$ is the temperature of the gas, the latter two of which are properties of the surrounding gas far from the accreting black hole.

Here we assumed the gas is composed of pure hydrogen and the adiabatic index is 5/3. The radiative efficiency $\epsilon$ determines the accretion luminosity for a given accretion rate,
\begin{equation}
L_{\rm bol} \equiv \epsilon \dot{M}_B c^2
.
\end{equation}
This allows us to define the dimensionless mass accretion rate and luminosity,
\begin{equation}
l
=
\frac{L_{\rm bol}}{L_{\rm Ed}},
\qquad
\dot{m}
=
\frac{\dot{M}_B}{\dot{M}_{\rm Ed}},
\end{equation}
where
\begin{equation}
L_{\rm Ed}
=
1.26\times 10^{38}\ \text{erg s}^{-1}\ 
\left(
\frac{M}{1\ M_\odot}
\right)
\end{equation}
and $\dot{M}_{\rm Ed} = L_{\rm Ed}/c^2$. This leads to the simple relation $l = \epsilon \dot{m}$.

Before discussing the emission of our system, we note that the accretion timescale for a solar-mass black hole surrounded by gas is
\begin{equation}
\tau_{B} = M_B/\dot{M}_B.
\end{equation}
For the parameters relevant to our examination of $G$ objects, this timescale exceeds the age of the Galaxy. This suggests that accretion of a $G$ object's atmosphere onto the solar-mass black hole at its center does not alter the system on the timescales we are considering.

It is notoriously difficult to estimate the accretion luminosity from first principles. Proper treatment requires numerically solving relativistic generalizations of the continuity equation, Euler's equation, and an entropy equation given by the thermodynamic identity. To solve these equations, one must also specify the equation of state for the gas and relevant heating and cooling mechanisms~\cite{Shapiro1973:AccretionI, Shapiro1973:AccretionII, Meszaros1975:Accretion, Blondin1986:HyperAccretion, Nobili1991:Accretion, Park:2001Preheated, Thorne1981:AccretionI, Flammang:1982AccretionII}. Instead of pursuing an imperfect numerical solution, we will instead parametrize the theoretical and experimental uncertainties in the efficiency through a free parameter $\eta$, which is defined such that $\epsilon = \eta\dot{m}$, as inspired by solutions obtained by Refs.~\cite{Shapiro1973:AccretionI, Shapiro1973:AccretionII, Meszaros1975:Accretion}.

\begin{figure}[htb]
    \includegraphics[width=0.95\linewidth]{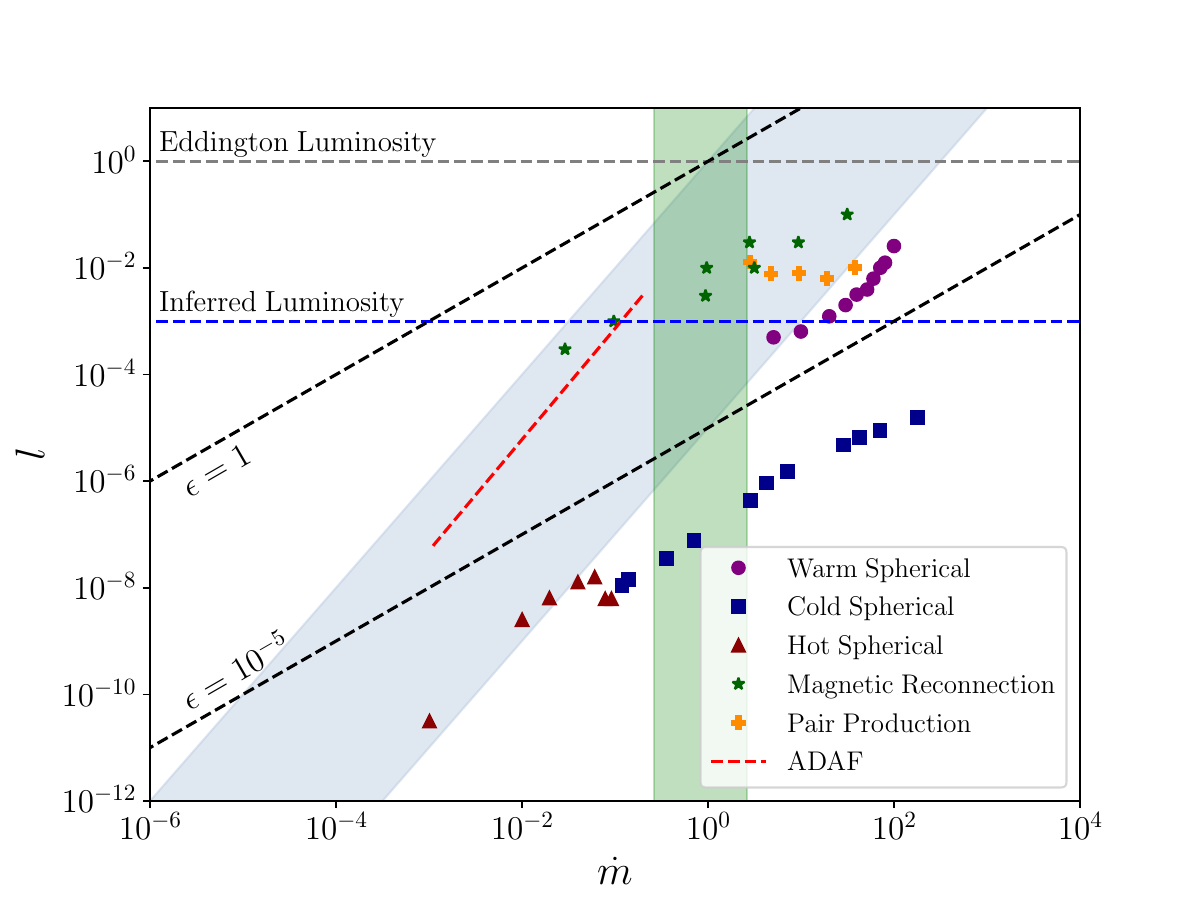}
    \caption{The dimensionless luminosity $l$ as a function of dimensionless accretion rate $\dot{m}$. The blue band represents varying values of $\eta$ between $10^{-5}$ and 1. The dashed blue line represents values of $l$ consistent with observation. The dashed gray line corresponds to the Eddington limit $l = 1$. The green region corresponds to values of $\dot{m}$ consistent with the assumption that $G$2 is an optically thin cloud of ionized gas~\cite{Gillessen:2012G2}. The data points illustrate various modes of accretion as calculated and categorized by  \cite{Park:2001Preheated}, which specifically focuses on advection-dominated accretion flows (ADAF).}
	\label{fig:AccrDiag}
\end{figure}

The blue region of Fig.~\ref{fig:AccrDiag} corresponds to varying $\eta$ between $10^{-5}$ and 1, with constant values of $\epsilon$ plotted for reference. In order to demonstrate that the values of $\eta$ we selected were appropriate, we also show data values computed and classified by Park and Ostriker~\cite{Park:2001Preheated}. Additionally, we also included the green shaded region, which corresponds to values of $\dot{m}$ consistent with the assumption that $G$2 is an optically thin cloud of ionized gas~\cite{Gillessen:2012G2}, namely, $T_{\infty} = 10^4$ K and $n_\infty = (0.2 $ - $ 2)\times 10^5$ g cm$^{-3}$. The intersection of the shaded regions suggests that emission of a solar-mass black hole with a $G$2-like atmosphere is consistent with the inferred luminosity of the class of $G$ objects. The required values of $\eta$ or $\epsilon$ are in-line with numerous accretion scenarios, which demonstrates the viability of our scenario.

Determining the spectrum of emission provided by the central black hole also requires a full numerical calculation. The atmosphere resulting from destructive PBH-NS conversion processes will contain many heavy elements that appear as dust~\cite{Fuller:2017uyd}. Depending on the density of the atmospheres surrounding a central solar-mass black hole, this dust may reprocess higher energy photons into the infrared spectrum. Regardless of these details, our examination of the total emitted luminosity demonstrates that a solar-mass black hole does not significantly under- or overproduce radiation.

As already mentioned, solar-mass PBHs, which could be a subdominant component of DM, would also be most commonly found in the GC. It is natural to consider that these larger PBHs on their own could attract gas and dust from the surrounding environment and potentially present themselves as $G$ objects. Given the uncertainly about the DM profile in the GC, the abundance of solar-mass PBHs in the central 0.04 pc is $\gtrsim \mathcal{O}(10)$,  which coincides with the observed population of $G$ objects. Unlike the sublunar-mass PBH-NS conversion, events however, solar-mass PBHs would have to acquire their atmosphere through accretion alone. Given the proximity to the central SMBH, these solar-mass PBHs would have large velocities that would suppress accretion and have drastic implications for the flow into the PBH core. This is also true for NSs or $\mathcal{O}(1 - 10) M_\odot$ black holes that might reside in the GC due to mass segregation~\cite{Miralda-Escude:2000kqv, Hopman:2006rt, Hopman:2006xn}, the latter of which are likely too massive to be consistent with the objects inferred at the center of the $G$ objects. A simple estimate of the Bondi-Hoyle accretion timescale for a solar-mass object within the GC indicates that these objects will not acquire enough mass within the lifetime of the galaxy to be consistent with the observed $G$ objects. However, direct interactions between stellar-mass black holes and stars may generate $G$ objectlike structures e.g.,~\cite{Rose:2021ftz,Kremer+22}.

\section{Discussion and Conclusion}

\label{sec:DiscConclusion}

In summary, the PBH interpretation of $G$ objects offers a unique insight into the nature of DM, the missing pulsar problem, and the origin of $G$ objects. This scenario provides a unique opportunity to investigate the existence of solar-mass black holes. The large abundance of DM in the GC allows this region to be a test bed for the PBH-DM hypothesis. If DM takes the form of sublunar PBHs, our calculations and others~\cite{Capela:2013CapNS, Kouvaris:2013kra, Takhistov:2017nmt, Fuller:2017uyd, Takhistov:2020vxs} suggest that conversion events  have the potential to be observed as kilonovae without a corresponding gravitational wave signal and are likely to only be observed in regions rich in DM. At the same time, the newly formed black holes will retain the velocities of their progenitor  and will be surrounded by initially neutron-rich material, which later acts as a thick atmosphere of dust and gas. Should $G$ objects be the result of NS conversions, they should only be observed where the DM density is significant, i.e., the GC. In combination with microlensing, gravitational waves, and other observational techniques, the existence of $G$ objectlike configurations allows for further investigation into the PBH-DM scenario.


\begin{acknowledgements}
We thank Briley Lewis whose writing inspired this work. Additionally, we thank Mark Morris for his valuable insights. Finally, we thank an anonymous referee for their insightful and helpful comments. M.M.F. and A.K. were  supported  by the U.S. Department of Energy (DOE) Grant No.  DE-SC0009937.  A.K.  was  also supported  by the Simons Foundation, by the World Premier International Research Center Initiative (WPI),  MEXT,  Japan,  by Japan Society for the Promotion of Science (JSPS) KAKENHI grant No. JP20H05853, and by the UC Southern California Hub with funding from the UC National Laboratories division of the University of California Office of the President. 
A.M.G. was supported by the National Science Foundation Grant  AST-1909554 and by the Heising-Simons Foundation, as well as Galactic Center Stars Society.  S.N. acknowledges the partial support from NASA ATP 80NSSC20K0505 and NSF AST-2206428 grants, and thanks Howard and Astrid Preston for their generous support. 
\end{acknowledgements}


\bibliography{biblio}

\end{document}